\documentstyle[epsf]{kluwer}
\input psfig.sty
\def\gapprox{\;\rlap{\lower 2.5pt            
 \hbox{$\sim$}}\raise 1.5pt\hbox{$>$}\;}       
\def\lapprox{\;\rlap{\lower 2.5pt            
 \hbox{$\sim$}}\raise 1.5pt\hbox{$<$}\;} 
\def\N{\,{\rm I\kern-.20em N}}
\begin{document}

\begin{article}
\begin{opening}
\title{CALLISTO - A New Concept for Solar Radio \\
Spectrometers}
\author{Arnold O. \surname{Benz}}
\author{Christian \surname{Monstein}}
\author{Hansueli \surname{Meyer}}

\institute{Institute of Astronomy, ETH Z\"urich, CH-8092 Zurich, Switzerland\\}
\runningtitle{A New Concept for Solar Radio Spectrometers}
\runningauthor{Arnold O. Benz et al.}

\date{Received:.......  ; accepted :....... }
\begin{abstract} 
A new radio spectrometer, CALLISTO, is presented. It is a dual-channel frequency-agile receiver based on commercially available consumer electronics. Its major characteristic is the low price for hardware and software, and the short assembly time, both two or more orders of magnitude below existing spectrometers. The instrument is sensitive at the physical limit and extremely stable. The total bandwidth is 825 MHz, and the width of individual channels is 300 kHz. A total of 1000 measurements can be made per second. The spectrometer is well suited for solar low-frequency radio observations pertinent to space weather research. Five instruments of the type were constructed until now and put into operation at several sites, including Bleien (Zurich) and NRAO (USA). First results in the 45 -- 870 MHz range are presented. Some of them were recorded in a preliminary setup during the time of high solar activity in October and November 2003. 

\end{abstract} 
\end{opening}
\section{Introduction}
The radio emission of solar flares comes in many varieties, probably reflecting different emission processes and physical origins. The emissions are distinct in their spectrum and temporal behavior. They can best be distinguished in spectrograms, as demonstrated already by Wild \& McCready (1950). In the meter wavelength range, 5 types of radio emissions are distinguished (review by McLean \& Labrum, 1985) and at decimeter wavelengths there is a similar number (G\"udel \& Benz, 1988; Isliker \& Benz, 1994). The radiations are caused coherently by electron beams, shocks, possibly trapped electrons, and high-frequency waves in the plasma. The incoherent gyrosynchrotron radiation is the only really broadband radio emission. As the radio emissions carry information on shocks, energetic electrons and their acceleration, but are still largely unknown, research in flare radio emissions will grow in the near future. In view of space weather implications, meter wave bursts have seen a renewed interest, particularly in combination with observations at other wavelengths. 

Spectrometers are necessary to identify the nature of coherent solar flare radio emission. Interferometers do not have sufficient frequency channels to replace spectrometers with high frequency resolution, as some decimeter bursts have bandwidths of less than one percent of the center frequency. Spectrometers will be needed in the future to complement interferometers such as the Frequency Agile Solar Radiotelescope (FASR, Bastian, 2004). 

The first radio spectrometers recorded optically the analog signal on film. The development of digitally recording spectrometers in the 1970s (Perrenoud, 1982) soon reached the physical limit of sensitivity for frequency-agile receivers given by the radiometer equation. As the quiet Sun is a strong background, increasing the antenna size helps only for field of views smaller than the full disk. In frequency-agile systems frequency channels are measured sequentially. The sensitivity can only be improved by multi-channel spectrometers, which measure multiple frequency channels simultaneously. Such instruments were developed earlier on, but had a limited number of channels and total bandwidth (Dumas, Caroubalos, \& Bougeret, 1982). This shortcoming was overcome with the development of acousto-optic receivers (Cole, Stewart, \& Milne, 1978). However, their limited dynamic range makes acousto-optic spectrometers vulnerable to terrestrial interference. For this reason, spectrometers of acousto-optic design have been operated successfully only at centimeter wavelengths where interference is less severe (Lecacheux et al., 1993; Benz, Messmer, \& Monstein, 2001). The more robust Fast Fourier Transform (FFT) technology is still in development. 

While the development of broadband, interference-resistant spectrometers is still a "desideratum", we are left at present with receivers operating according to the frequency-agile principle. There is some development in recent years with regard to broader bandwidth, circular polarization, and better software for operation, calibration and data analysis. Frequency-agile or swept-frequency spectrometers are currently in operation in several places, including China (Xu et al., 2003), Hiraiso (Kondo et al. 1995), Ondrejov (Jiricka et al., 1993), Sao Paolo (Sawant et al., 2001), and Tremsdorf (Mann et al., 1992). ETH Zurich also operates broadband spectrometers since many years. The Ikarus-Phoenix line of digital frequency-agile receivers has seen three decades of steady, but costly developments in hardware and software (Benz et al., 1991). 

Today the situation has changed dramatically. The hardware and some software have become commercially available on the consumer electronics market. Thus radio spectrometers can be built cheaply and in great numbers. They are small and transportable. This opens new possibilities, in particular concerning the ever increasing interference. Identical spectrometers could be placed far apart to observe different interference-free windows. Alternatively, a spectrometer can be brought temporarily into a remote region with less interference. 

Here we describe the new radio spectrometer CALLISTO (Compact Astronomical Low-frequency, Low-cost Instrument for Spectroscopy in Transportable Observatories) built according to these principles. Its main focus is solar flare radio observation.

\section{Concept of CALLISTO}

\subsection{Hardware}

CALLISTO is a frequency-agile spectrometer that is easily transportable and can be used in many observatories. It has been operated with a 5 m parabolic reflector (from 80 MHz to 800 MHz) near the Institute in Zurich and, more recently, with a log-periodic yagi antenna (gain 6 db) in the full frequency range at the Bleien Radio Observatory about 50 km west of Zurich.

\begin{figure}
\begin{center}
\leavevmode
\mbox{\hspace{-0.2cm}\epsfxsize=11.2cm
\epsffile{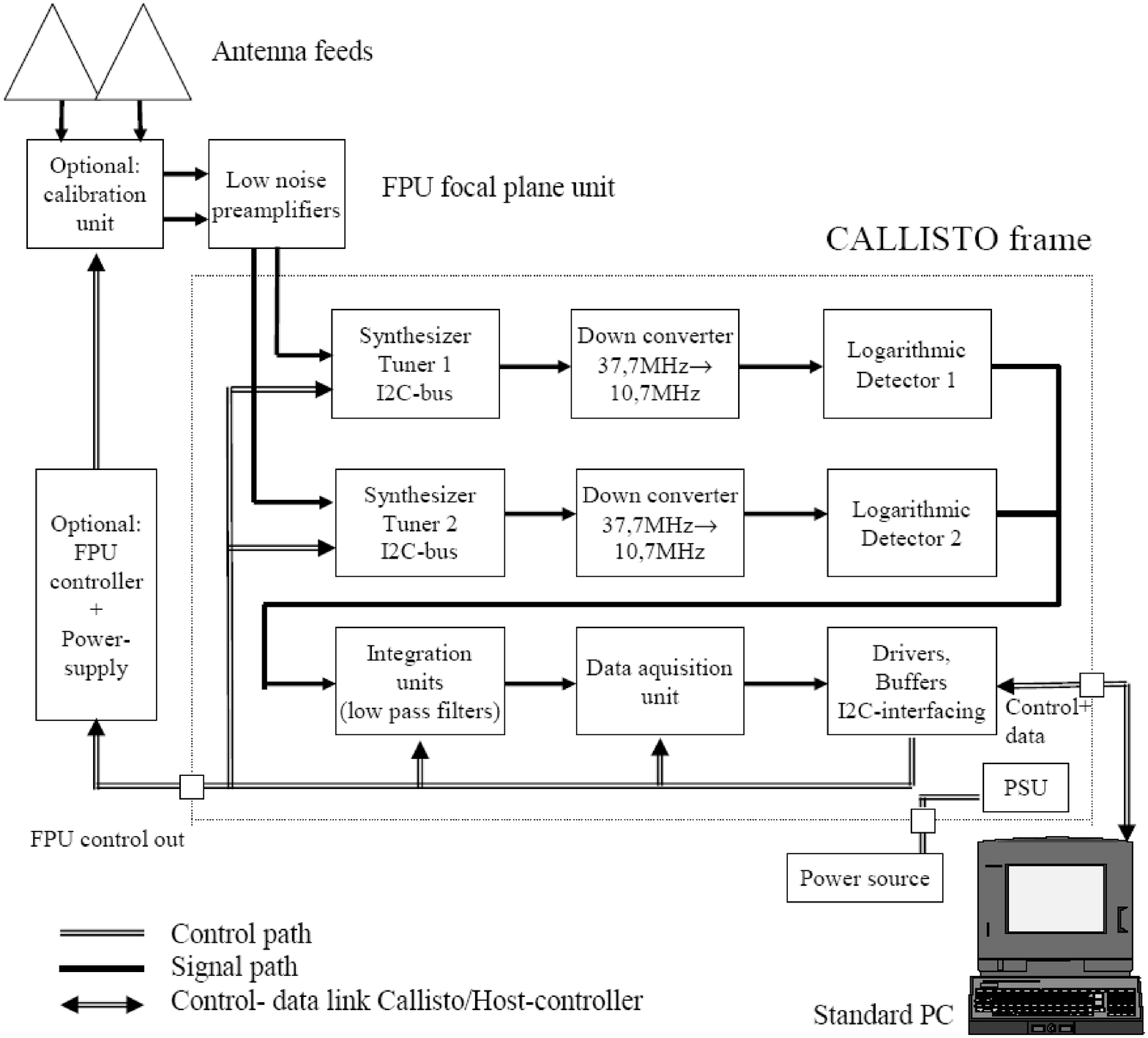}}
\end{center}
\caption[]{Basic design of the radio spectrometer CALLISTO for solar flare observations. Two receivers (top and middle row) operate in phase.}
\label{schema}
\end{figure}

The basic diagram is shown schematically in Fig. \ref{schema}. The signals from two feeds (e.g. two modes of polarization) are fed into two receivers (top row and middle row). They convert to a first intermediate frequency of 37.7 MHz by two local oscillators (digital tuners) controlled by the online PC. Then the signal is down converted to 10.7 MHz for filtering and amplification, detected by a logarithmic device, and low pass filtered. The logarithmic domain is more than 45 db, where the deviation from logarithmic is less than 0.3 db. Data acquisition for both receivers and the interface to the PC are on a separate board. The two receivers work in phase. The measurements are made in a two step process. In the first step a receiver is tuned to a frequency, in the second step the signal is measured. The two steps take equal time of the order of 0.5 ms. The two receivers can also be configured to measure the same polarization and to alternate: while one is measuring, the other is tuned to a new frequency. 

The hardware is depicted in Fig. \ref{hardware}. Its total cost is less than one percent of the components for Phoenix-2 (built in late 1980s). Also, the CALLISTO spectrometer can be rebuilt from commercially available consumer hardware in about two orders of magnitude less time.

The timing of CALLISTO is controlled by a GPS clock. Thus the relative timing is accurate to within less than one millisecond, whereas the absolute time is uncertain to within a few milliseconds due to internal delays.

The total frequency range is from 45 to 870 MHz. An individual channel has 300 kHz bandwidth and can be tuned by the controlling software in steps of 62.5 kHz. The narrow channel width and accurate positioning are essential to avoid known bands of interference. The number of channels per observing program is limited between 1 and 500, and a total of up to 1000 measurements can be made per second. For a typical number of 200 channels, the sampling time per channel is 0.2 seconds. The instrument can easily be duplicated for a larger number of channels or can be used at higher frequencies by down-converting to the lower frequency range mentioned above. More information is available under http://www.astro.phys.ethz.ch/instrument/ callisto/callisto.html.

\subsection{Software}

The software is distributed on a RISC processor ATmega16 and a standard PC or laptop. On the RISC, the driver, buffer and interfacing software is programmed in C$^{++}$, using an interrupt-driven state machine concept. The host software on the PC is also in C$^{++}$ and operates under Windows 2000 and XP.

The relevant parameters are locally stored in a text file which can be easily adapted to other observing configurations. Additional RS232 ports are preconfigured to communicate with an extended GPS system and external temperature and humidity sensors. It is also possible to control CALLISTO via Internet, using an RS232 network adapter. A file-controlled scheduler starts and stops measurements in relation to local PC time or UT. The scheduler is repeated every day automatically and can be changed on-line and remotely.

\section{Tests, Calibration and Data Analysis}

The Allan variance was measured using a 50 Ohm resistor at a controlled temperature. The minimum Allan variance was found at 4000 seconds.

\begin{figure}
\begin{center}
\leavevmode
\mbox{\hspace{-0.2cm}\epsfxsize=10.2cm
\epsffile{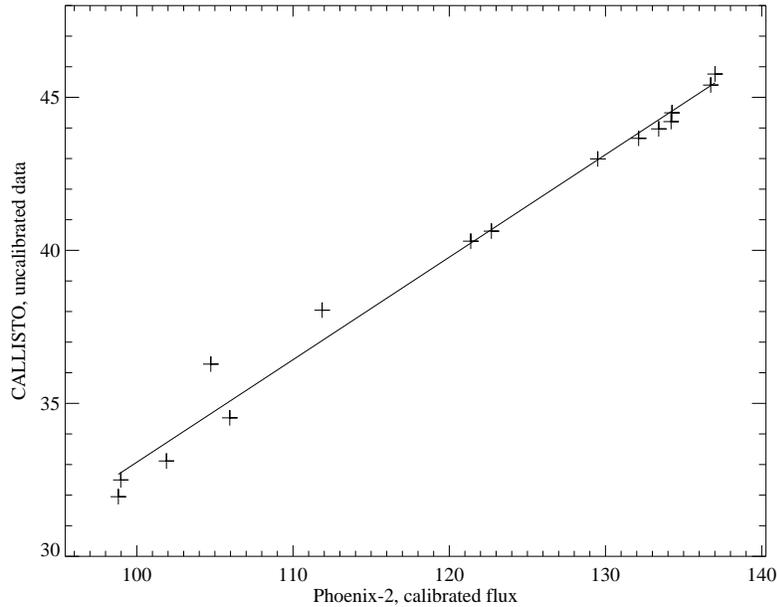}}
\end{center}
\vskip-0.1cm
\caption[]{The CALLISTO raw data vs Phoenix-2 calibrated data in logarithmic compression. The displayed range of 45 units for Phoenix-2 amounts to a factor 10 in solar flux. The measurements have been made simultaneously at the same frequency of 408 MHz. The linear regression line is also shown.}
\label{calibration}
\end{figure}

Until now we have calibrated CALLISTO by comparison with the Phoenix-2 spectrometer (Messmer et al., 1998). Figure \ref{calibration} displays uncalibrated CALLISTO data vs. calibrated Phoenix-2 measurements during a radio burst. In lin-log representation, the relation is linear. Therefore the uncalibrated CALLISTO data must be very close to logarithmic. This is not surprising as its detectors are indeed specified to follow the logarithmic curve within a few percent. At frequencies below the range of Phoenix-2, the calibration is based on man-made sources, such as satellites transmitters with well-known power and ground-based test emitters.

The output, logarithmically compressed by the detector, has 10 bits, of which the 8 most significant ones are stored. The data transfer is initiated every night by the server in Zurich. The data are calibrated, quick-look spectrograms are put on the web, and the calibrated data is archived in the same format as Phoenix-2 in the HESSI Experimental Data Center (HEDC) operated by the Institute of Astronomy.

Since the archived data formats are identical, the same programs for scientific data analysis can be used as for Phoenix-2 data. These are the IDL-based package "Ragview", available on the SolarSoftWare (SSW) system, the RHESSI related program "spectroplot" (also SolarSoftWare), and a Java script called "RappViewer".

\begin{figure}
\begin{center}
\leavevmode
\vskip-1.5cm
\mbox{\hspace{-0.2cm}\epsfxsize=12.2cm
\epsffile{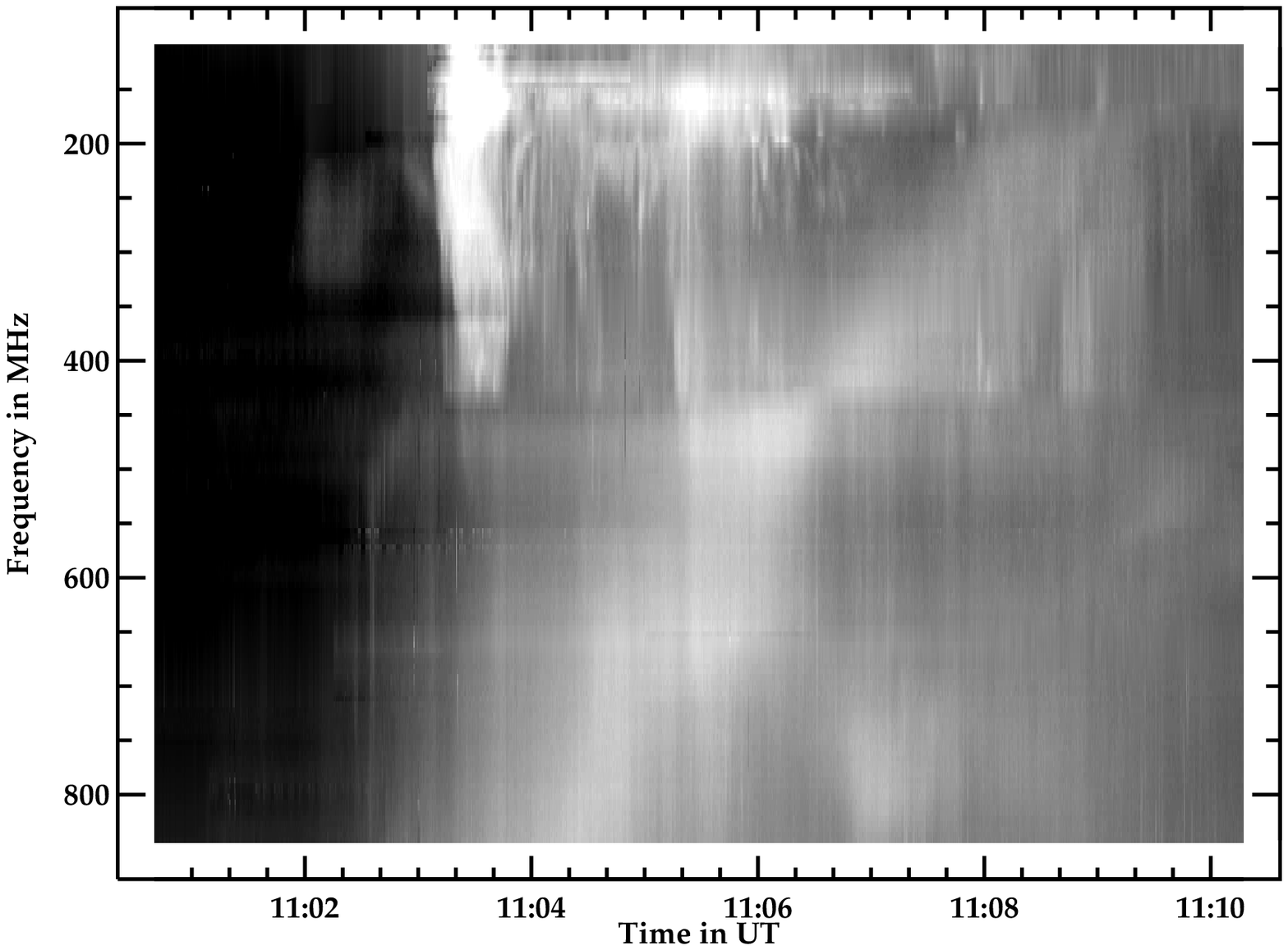}}
\end{center}
\vskip-0.5cm
\caption[]{The X17.2 flare of 2003 October 28 was observed by CALLISTO in a temporary setup. The image is cleaned from terrestrial interference.}
\label{12X}
\end{figure}

\section{First Results}

CALLISTO has just been completed when the Sun had its period of great activity in October and November 2003. The new radio spectrometer was temporarily installed with a 5 m parabolic reflector dish. Its feed is sensitive down to 160 MHz, but very limited below. In this preliminary setup, CALLISTO observed serendipitously several X-class flares. Figure 4 presents the largest flare observed so far. The first group of radio bursts at 80 -- 500 MHz coincides with very intense peak of centimeter emission at 15.4 GHz reported by Solar and Geophysical Data and with the peak of the GOES derivative. It consists of at least 4 structures with FWHP duration of 5.1 s drifting to higher frequency at a rate of +17 MHz/s at 250 MHz. Below about 200 MHz the drift is to lower frequency with about the same rate. The flux reaches values of up to 7$\cdot 10^4$ sfu between 210 -- 290 MHz. As the Phoenix-2 spectrometer saturates below $10^4$ sfu, this is the largest flux ever measured by the Zurich spectrometers. The slow drift rate and high flux value are untypical for type III bursts, but resemble more the drifting patches observed at decimeter wavelengths (Smith \& Benz, 1991). The first patches are followed by similar structures at lower flux levels for several minutes. At higher frequencies, one minute later, a second emission appears at 850 MHz, drifting with -2.15 MHz/s at 380 MHz across the whole frequency band of high sensitivity to about 153 MHz. The drift rate is in the range of meter-wave type II bursts, but the peak flux of some 4$\cdot 10^4$ sfu at 400 MHz and the bandwidth of some 200 MHz at 400 MHz are very unusual. As with most large flares, there was a large CME associated.

\begin{figure}
\begin{center}
\leavevmode
\vskip-0.5cm
\mbox{\hspace{-0.02cm}\epsfxsize=12.2cm
\epsffile{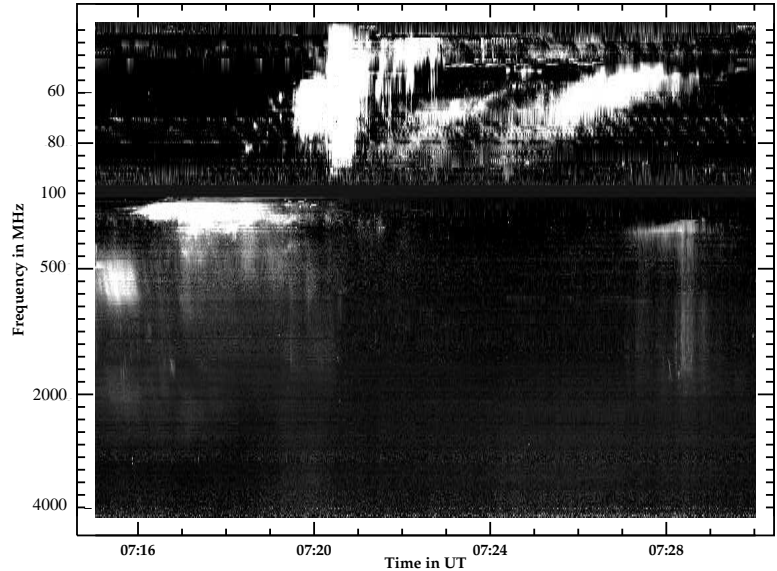}}
\end{center}
\caption[]{The GOES-class C3.1 flare of 2004 April 27 was observed with CALLISTO at Bleien Observatory from 100 MHz down to the lower frequency limit of 45 MHz. This setup complements the observing range of Phoenix-2 from 100 MHz - 4000 MHz, shown in the lower part of the figure, separated by a black region. }
\label{April272004}
\end{figure}

A more typical spectrogram is shown in Fig. 5. The CALLISTO spectrometer is used to complement Phoenix-2 (100 -- 4000 MHz) at the lower meter-wave frequencies (45 -- 100 MHz). A first group of meter-wave type III bursts is followed by two type II bursts. A third type II-like structure appears later in the Phoenix-2 spectrogram at 300 MHz. Other emissions in the Phoenix-2 part of the spectrogram include decimetric emissions such as drifting patches at 600 MHz early in the flare and type III bursts at 300 MHz.

\section{Conclusions}
We have presented the new solar radio spectrometer CALLISTO now in operation at the ETH Zurich radio observatory in Bleien (Switzerland). In the 45 -- 870 MHz range CALLISTO surveys a spectral region of renewed interest in view of space weather implications. It is a very cheap instrument suitable for surveying the radio spectrum at low frequencies. Copies can easily be made and distributed to other locations. Its principle aims are low-frequency observations for space weather investigations, complementary observations for interferometers, and routine observations for which it could eventually replace aging spectrometers. Five identical instruments have been realized until today. Two are currently operating near Zurich to complement the existing Phoenix-2 spectrometer below its low-frequency limit and in the meter wave band (150 -- 400 MHz), where the small channel width of CALLISTO is exploited to observe between interferences (http://www.astro.phys.ethz.ch/rapp/). A third instrument is being installed at NRAO in Greenbank (USA) to complement the Green Bank Solar Radio Burst Spectrometer in the range 70 to 870 MHz. Two instruments will be used in temporary set-ups at remote locations free of interference. 

\begin{acknowledgements}
Pascal Behm and Ivo Zamora have contributed to the construction, testing and duplication of CALLISTO. We acknowledge the diploma work of Hubert Perret that includes the calibration of CALLISTO and Fig. 3. We thank also Pascal Saint-Hilaire, Christina P\"opper and Peter Steiner for software support, as well as Frieder Aebersold for providing mechanical components and help. The construction of the CALLISTO spectrometer is financed by the Swiss National Science Foundation (grant nr. 20-67995.02).
\end{acknowledgements}


\end{article}
\end{document}